\documentclass[jcp,preprint,showkeys,nofootinbib]{revtex4-1}

\usepackage{amsmath,amssymb,amscd,enumerate, mathrsfs, latexsym,ulem}
\usepackage{graphicx}
\usepackage{xcolor}
\usepackage{hyperref}

\begin{document}
\title{Comment on ``On Uniqueness of SDE Decomposition in A-type Stochastic Integration''[arXiv:1603.07927v1]}
\author{Peijie Zhou}
\email{cliffzhou@pku.edu.cn}
\author{Tiejun Li}
\email {tieli@pku.edu.cn}. 
\thanks{Corresponding author.}
\affiliation{LMAM and School of Mathematical Sciences, Peking University,  Beijing 100871, China}
\begin{abstract}
The uniqueness issue of SDE decomposition theory proposed by Ao and his co-workers has recently been discussed. A comprehensive study to investigate connections among different landscape theories [J. Chem. Phys. 144, 094109 (2016)] has pointed out that the decomposition is generally not unique, while Ao  et al. (arXiv:1603.07927v1) argues that such conclusions are ``incorrect'' because of the missing boundary conditions. In this comment, we will combine literatures research and concrete examples to show that the concrete and effective boundary conditions have not been proposed to guarantee the uniqueness, hence the arguments in [arXiv:1603.07927v1] are not sufficient. Moreover, we show that the ``uniqueness'' of the O-U process decomposition referred by YTA paper is unable to serve as a counterexample to ZL's result since additional assumptions have been made implicitly beyond the original SDE decomposition framework, which cannot be applied to general nonlinear cases. Some other issues such as the failure of gradient expansion method will also be discussed. Our demonstration contributes to better understanding of the relevant papers as well as the SDE decomposition theory.
\end{abstract}

\maketitle
\section{Introduction}
As the energy landscape attracts much attention in biophysics community in recent years, different proposals have been put forward in recent literatures. The paper ``Construction of the landscape for multi-stable systems: Potential landscape, quasi-potential, A-type integral and beyond" \cite{ZL} (abbreviated as ZL paper in later text) is intended to clarify the relationship between different proposals. The ZL paper, to our knowledge, for the first time provides rigorous mathematical results for the general existence of SDE decomposition proposed by Ao and his collaborators \cite{Ao2004} by relating it to the famous Freidlin-Wentzell theory, and gives theorems as well as concrete counter-examples to deny the uniqueness of the decomposition in high-dimensional nonlinear case.  

Recently, Yuan, Tang and Ao \cite{yta} (abbreviated as YTA paper in later text) assert that the non-uniqueness result about the SDE decomposition in ZL paper was ``incorrect'' because the ``original definition'' of the SDE decomposition already contains the boundary conditions to ensure the uniqueness. In this comment, we will combine both literatures research and concrete examples to illustrate the flaws of such conclusions: The previous literature on the SDE decomposition {\it never} gave a concrete and effective condition to guarantee the uniqueness of the decomposition, and the seemingly more explicit condition appeared in YTA paper is invalid to eliminate the degrees of freedom found by ZL paper. A detailed discussion about the O-U process will then be provided, indicating that when proving the uniqueness of the decomposition for linear cases \cite{pnas},  some further conditions are implicitly imposed. The original and general proposal on SDE decomposition is not consistent with such conditions, which renders the ``uniqueness result" on O-U processes invalid to challenge ZL's non-uniqueness proof. We will also present the failure of the so-called ``gradient expansion method'', which has been used as an argument to establish the uniqueness of the decomposition \cite{Ao2005, Ao2009}. Some other comments in YTA paper will be responded as well.

The comment is organized as follows. Previous literatures related to the boundary condition and uniqueness issue of SDE decomposition will be explored in Section 2. Then the concrete example that appeared in the Supplementary Material of ZL paper will be revisited in Section 3 to show the invalidity of the boundary condition raised in YTA paper, as well as the failure of the gradient expansion method. We will respond to some other points raised in YTA paper in Section 4.

\section{Boundary Condition and Uniqueness in Ao's theory}
\subsection{Survey on Previous Literatures: An Already-Existed Boundary Condition?}
One of the main point of the YTA paper \cite{yta} is that Ao and his collaborators have already proposed a boundary condition (4 lines below Eq. (1b) in page 2) to ensure the uniqueness in their previous works. To check this statement, we made an incomplete yet careful search on the literature to explore the role of boundary condition as well as the uniqueness statement of SDE decomposition in Ao's theory. 

\begin{itemize}
			\item  In their 2004 paper \cite{Ao2004}, it is stated that  ``\textit{We prove the existence and uniqueness of the gauged $\phi$- decomposition from equation (1) to (4) by an explicit construction}" and ``\textit{Equations (10) and (11) give the needed $n \times n$ conditions to completely determine G. Here we give a solution of G as an iteration in gradient expansion}". 
			
			To our knowledge, the 2004 paper is the first systematic presentation to the general framework (and the ``original definition'', as claimed in \cite{yta}) of Ao's SDE decomposition theory,  and the ZL's introduction and discussion on SDE decomposition in \cite{ZL} is mainly based on this paper. It is very unfortunate that we do not find {\it any explicit statement relevant to boundary conditions} in this paper. Even so, {\it the paper already stated their ``uniqueness" result} in this stage. The existence and uniqueness of the solution for a nonlinear PDE system is not trivial in general. One serious scientist can speculate, or expect the well-posedness of the problem, but should be open to the rigorous mathematical studies on this. Physicists can construct the objects in their mind, but not refuse the possible defects on their theory  until final justification. 
			
			It should also be noted the so-called ``gradient expansion method'' was proposed in this paper \cite{Ao2004}. In some later literatures, it has been claimed that such method will imply certain ``boundary conditions'' and give the unique solution. However, such claims are problematic. Firstly, gradient expansion method is just a unrigorous formula to give one special solution of the PDE systems, it can not replace rigorous mathematical results of uniqueness. Secondly, such expansion has nothing to do with the ``boundary condition'', at least not in the common sense like Dirichlet or Neumann boundary condition for a PDE. What is more, {\it the gradient expansion may not work (i.e. the infinite sum might not converge) for even  very simple problems.} A relevant example will be provided in Section 3. Therefore, it is not convincing to equate the ``gradient expansion method'' to the uniqueness issue of SDE decomposition.
			
			This supports the starting point of ZL paper on SDE decomposition, which is exactly based  on the original definition of the theory where no boundary conditions are attached.

\item In their 2005 paper \cite{Ao2005}, the ``boundary condition" was mentioned for the first time ``\textit{Hence, with an appropriate boundary condition, $G$ can be found, so will be $A$, $T$ , and $\psi$}". 

But the conditions are just used to ``find'' the solution, no uniqueness result for the PDE system is discussed, neither did they specify what is ``an appropriate boundary condition".
			
			\item In their 2008 paper \cite{Ao2008}, they said ``\textit{The boundary condition in solving Eq. is implied by the requirement that the fixed points of $f$ should coincide with the extremals of the potential function $\phi$}". 
			
			Such condition does nothing to guarantee the uniqueness of the PDE (all the counterexamples against the uniqueness of SDE decomposition provided below will satisfy such condition) and it is also {\it not a workable and verifiable condition} when one solves the PDE system (Eq. (1) in the YTA paper).
			
			\item In their 2009 review paper \cite{Ao2009}, they said ``\textit{Thus, Eq.(4) is precisely the potential function condition. It can be found via the integration over Eq.(5), independent of the integration routes connecting initial and final points. All $A$, $T$, $Q$, as well as the potential function $\psi(q)$ are uniquely determined by the diffusion matrix and the deterministic force $f(q)$}". 
			
			Again, no boundary condition for PDE is mentioned, while the authors still  claimed the uniqueness (their ``uniqueness'' comes from the ``gradient expansion'' method. But as we mentioned before, this approach fails in many high-dimensional nonlinear cases and it is also not sure whether it depends on the choice of the initial value. It is just a intuitive treatment without mathematical proof on its convergence and therefore does not help to establish the theoretical uniqueness of the PDE).
			
			\item  In their 2012 paper \cite{Ao2012}, it is claimed ``\textit{In principle, the potential function $\phi(q)$ can be derived analytically by solving the $n(n-1)/2$ partial differential equations (under proper boundary conditions) of equation (8)}" without pointing out concretely what is the ``proper boundary conditions". And the uniqueness issue is also not mentioned in the paper.

\end{itemize}

To summarize, before ZL's paper, previous literatures on SDE decomposition {\it never} gave a concrete, effective and explicitly verifiable condition to show the uniqueness (even existence) of the decomposition and A-type integral. They usually use the words like {\it ``proper"} or some conditions that are difficult to be verified (to our knowledge these conditions cannot eliminate the possibilities of non-uniqueness as pointed out via ZL's example in their supplementary material), and they even reach the uniqueness conclusion without any reference to the boundary conditions. Moreover, we would like to stress that the so-called ``gradient expansion'' is just a numerical strategy whose rigorous convergence results are not yet established, which of course, can not serve as the ``proof'' to the uniqueness of the PDE systems.

Meanwhile, in the recent YTA paper after ZL's result, the authors modified their conclusions as ``\textit{the existence and uniqueness of the solution is guaranteed at least locally}'' with one class of boundary condition that ``\textit{near fixed points every component of Q is a smooth function of state variable}''. Although the conclusions seem to be weaker and the boundary conditions appear to be more concrete,  unfortunately,  they still cannot rule out the counter-examples raised in ZL paper (see also the next section), because all the components of possible $Q(x)$ constructed in ZL paper are the smooth functions of state variables.

\subsection{``Uniqueness of the Decomposition" for O-U Processes: A Counter-Example?}
\label{sec:OU}
It might be very misleading and deceptive  to use the unique decomposition for O-U processes in \cite{pnas} as a ``counter-example'' to challenge ZL's non-uniqueness result. The first is the logic issue: Even if the uniqueness of the decomposition for special O-U processes is true, it is not sufficient to deny the non-uniqueness proof in \cite{ZL} for general SDEs, which is mathematically rigorous. Furthermore, after careful comparison and analysis, we will soon realize that a strong condition is implicitly imposed in \cite{pnas} to ensure the uniqueness, which is neither contained nor consistent with the SDE decomposition framework for arbitrary SDEs. Once such conditions are removed, the degree of freedom will naturally arise, which is shown in ZL's proof.

Let us begin by reviewing the arguments in \cite{pnas} and put them under the general framework of SDE decomposition. For O-U processes, one still needs to solve the same PDE system as presented in \cite{Ao2004, ZL,yta}
\begin{align}
G(x)+G(x)^{\tau} = 2D, \label{eqn:symm}\\
\nabla\times[G^{-1}(x)Bx] = 0,\label{eqn:curl}
\end{align}
where $B$ and $D$ are the constant drift and diffusion matrices in the considered O-U process. To determine the existence and uniqueness of such system, what the authors have done in \cite{pnas} actually utilized a key implicit assumption beyond the general framework proposed in \cite{Ao2004}: \textit{$G(x)$ is a constant matrix independent of the state variable $x$}. Only in this case Eq.~(\ref{eqn:curl}) can be rewritten as 
\begin{equation}
BG^{\tau}-GB^{\tau} = 0,
\label{eqn:matrixeqn}
\end{equation}
otherwise the derivatives of $G^{-1}(x)$ will contribute additional terms. Combining Eqs.~(\ref{eqn:symm}) and (\ref{eqn:matrixeqn}), the authors of \cite{pnas} then turn to discuss about the existence and uniqueness of the linear system
\begin{equation}
BQ+QB^{\tau} = BD-DB^{\tau},
\label{eqn:pnaslinear}
\end{equation}
where $Q$ is the anti-symmetric part of the matrix $G$.

In fact, if we stick to the original definition of the SDE decomposition \cite{Ao2004} as cited by ZL paper and remove the implicit condition on $G(x)$,  it is very easy to discover the degree of freedom for the decomposition of O-U processes, therefore showing no contradiction with ZL's non-uniqueness proof. For instance, let us consider a simple O-U process:
\begin{equation}
\begin{cases}
dX_t=-X_t dt+\sqrt{2\varepsilon}dW^{1}_{t}, \\
dY_t=-Y_t dt+\sqrt{2\varepsilon}dW^{2}_{t}, \\
dZ_t=-Z_t dt+\sqrt{2\varepsilon}dW^{3}_{t},
\end{cases}
\label{exampleou:sde}
\end{equation}
where $W^{j}_{t}~(j=1,2,3)$ are independent Brownian motions. The quasi-potential is  $\phi^{QP}=(x^2+y^2+z^2)/2$ and following ZL's result, we can construct $Q(x)$ with one degree of freedom, satisfying all the requirements proposed in general SDE decomposition framework \cite{Ao2004}

\begin{equation}
Q_{\lambda}(x,y,z)=\left(
\begin{matrix}
 0 & \lambda z & -\lambda y  \\
-\lambda z & 0 & \lambda x \\ 
\lambda y & -\lambda x & 0  \\
\end{matrix}
\right),
\label{eqn:ouQ}
\end{equation}
where $\lambda(x,y,z)$ is an arbitrary smooth function of $x,y$ and $z$. Only when we impose further restrictions as introduced in \cite{pnas}, can we determine that $\lambda(x,y,z)=0$ and obtain the unique decomposition $Q=0$, which can be also obtained from Eq.~(\ref{eqn:pnaslinear}). In this example, the constant $Q$ assumption is indeed a reasonable restriction from physical point of view which reflects the detailed-balance nature of the system. But for the cases when $B$ is not symmetric, which results in non-equilibrium dynamics without detailed-balance, the rationale to hold $Q(x)$ constant is questionable. 

As have been discussed above, the real problem for SDE decomposition proposed in \cite{Ao2004} is that, there has never been an available condition to ensure the uniqueness of decomposition in general cases as explicitly stated and rigorously proved as in the O-U process. Previous literatures by Ao and his collaborators have not even recognized this problem before ZL's paper. It is unfortunate that the requirement that \textit{$G$ is independent of $x$} can not be generalized to arbitrary SDEs other than O-U processes. For example, assume both $D$ and $Q$ are constant matrices while $b(x)$ is nonlinear, then from the fact that $b(x)=-(D+Q)\nabla\phi(x),\nabla\cdot(Q\nabla\phi)=\sum\limits_{i}\partial_{i}(\sum\limits_{k}q_{ik}\partial_{k}\phi)=\sum\limits_{i}\sum\limits_k{}q_{ik}\partial^2_{ik}\phi=0$ and $\nabla\phi^{\tau}Q\nabla\phi=0$, we know the SDE
\begin{equation*}
dX_{t}=b(X_{t})dt+\sigma dW_{t},\quad \sigma\sigma^{\tau}=2\epsilon D 
\end{equation*}
can be expressed as the form
\begin{equation*}
dX_{t}=g(X_{t})dt-D\nabla\phi(X_{t})+\sigma dW_{t},\quad \sigma\sigma^{\tau}=2\epsilon D,\nabla\cdot g =0, \langle\nabla\phi,g\rangle=0. 
\end{equation*}
This class of stochastic processes has important thermodynamical meanings \cite{qian}, while certainly not any SDE can be written in this form (the SDE considered in next section will serve as a counter-example). Hence, the restrictions on $G(x)$ in \cite{pnas} cannot be generalized.

Thus, in contrast to serving as a counter-example, the ``unique decompostion" of O-U process with additional assumptions instead strongly supports the conclusion in ZL's paper that \textit{under the current framework in the proposal \cite{Ao2004} the decomposition is generally not unique when the dimension of SDEs (1) is bigger than or equal to 3} (first paragraph in the right column of page 2 in ZL paper \cite{ZL}), and it also points out the significance of ZL's question \textit{Could there exist any other restrictions on S and A besides PDE systems which helps determine the decomposed process uniquely?} (second  paragraph in the left column of page 13 in ZL paper) The restriction that \textit{G does not depend on state variable $x$}  raised in \cite{pnas} is neither contained nor consistent with the general SDE decomposition framework in nonlinear cases, and the removement of such restriction in O-U processes will naturally reach the non-uniqueness result in ZL paper. We should also note that such restriction has nothing to do with the boundary condition. It is not the ``proper boundary condition'' that help uniquely determine the decomposition for O-U processes.

\section{A Concrete Example}
We will review the concrete example constructed in the Supplementary Material of ZL paper to further support our previous discussions. Specifically we will show that the so-called ``boundary conditions'' proposed in YTA paper are invalid to rule out the degree of freedom for SDE decomposition discovered in ZL paper, and the gradient expansion method proposed in \cite{Ao2004} fails in this simple example, render it useless to serve as a rescue to guarantee the uniqueness of the decomposition. 

Consider the SDE system
\begin{equation}
\begin{cases}
dX_t=(-X_t+Y_{t}^{2})dt+\sqrt{2\varepsilon}dW^{1}_{t}, \\
dY_t=(-Y_t-X_{t}Y_{t})dt+\sqrt{2\varepsilon}dW^{2}_{t}, \\
dZ_t=-Z_tdt+\sqrt{2\varepsilon}dW^{3}_{t},
\end{cases}
\label{example3d:sde}
\end{equation}
where $W^{j}_{t}~(j=1,2,3)$ are independent Brownian motions. The quasi-potential is readily solved from HJE with $\nabla\phi^{QP}=(x,y,z)^{T}$.

From the reconstruction procedure in ZL paper,  we can obtain
\begin{equation}
Q_{\lambda}(x,y,z)=\left(
\begin{matrix}
 0 & -y+\lambda z & -\lambda y  \\
y-\lambda z & 0 & \lambda x \\ 
\lambda y & -\lambda x & 0  \\
\end{matrix}
\right),
\label{eqn:Q}
\end{equation}
where $\lambda(x,y,z)$ is an arbitrary smooth function of $x,y$ and $z$. We will comment on YTA paper based on this example from two aspects.
\begin{itemize}
\item Ineffectiveness of boundary conditions in YTA paper and previous literatures.

By direct verification, we will show that the so-called ``boundary conditions'' proposed in the recent YTA paper as well as the previous literatures are invalid to exclude the degree of freedom discovered in ZL paper.

In the YTA paper, a class of boundary conditions are stated as ``near fixed points every component of $Q$ is a smooth function of state variable''. The elements of the non-unique $Q$ with one degree of freedom constructed in Eq.(\ref{eqn:Q}) are all smooth functions over the whole space. Hence, this ``boundary condition'' does not work.

In the 2008 paper \cite{Ao2008}, the ``boundary condition'' is thought to be implied by the requirement ``that the fixed points of $f$ should coincide with the extremals of the potential function $\phi$''. This condition has nothing to do with the matrix $Q$, therefore does not help to uniquely determine the decomposition.

\item Failure of the gradient expansion method.

The gradient expansion method can be written as an iteration scheme \cite{Ao2004,Ao2005}
\begin{align}
&G=D+Q, \notag\\
&Q = \lim_{j\to\infty}\Delta G_{j}, \notag\\
&\Delta G_{j} = \sum_{l=1}^{\infty}(-1)^{l}[(F^{\tau})^{l}\tilde{D}_{j}F^{-l}+F^{-l}\tilde{D}_{j}(F^{\tau})^{l}], \notag\\
&\tilde{D}_{0} = FD-DF^{\tau}, \notag\\
&\tilde{D}_{j\ge 1} = (D+\Delta G_{j-1})\{\partial\times[D^{-1}+\Delta G_{j-1}^{-1}]\}b(D-\Delta G_{j-1}).\label{eq:GEM}
\end{align}
where $F(x)$ is the Jacobian of $b(x)$. First of all, we need to note that such formulation is neither serious nor rigorous. For instance, when computing $D_{1}$, we need the inverse of the anti-symmetric matrix $\Delta G_{0}$, which is ill-defined when the dimension $n$ is an odd number. We doubt that Eq.~\eqref{eq:GEM} is a wrong formula, and a reasonable speculation is that $(D+\Delta G_{j-1})^{-1}$ is mistakenly written as $(D^{-1}+\Delta G_{j-1}^{-1})$. We suggest Ao and his collaborators to admit and clarify such problems. Secondly, there is no guarantee that the iteration will eventually converge. 

In this example, we have $D=I$ and 
\begin{equation*}
F(x,y,z)=\left(
\begin{matrix}
-1 & 2y & 0  \\
-y & -1-x & 0 \\ 
0 & 0 & -1  \\
\end{matrix}
\right).
\end{equation*}

When we attempt to utilize gradient expansion formula to compute $\Delta G_{0}$, the divergence of infinite sum $\sum\limits_{l=1}^{\infty}(-1)^{l}[(F^{\tau})^{l}\tilde{D}_{0}F^{-l}+F^{-l}\tilde{D}_{0}(F^{\tau})^{l}]$ will be encountered. 
Fig.~(\ref{fig:diverge}) shows the value of the $(2,1)$ element of the partial sum $\sum\limits_{l=1}^{L}(-1)^{l}[(F^{\tau})^{l}\tilde{D}_{0}F^{-l}+F^{-l}\tilde{D}_{0}(F^{\tau})^{l}]$
at point $(x,y,z)^{\tau}=(0.1,0.1,0.1)^{\tau}$ with different choice of the sum upper bound $L$, indicating that we cannot get a well-defined $\Delta G_{0}(0.1,0.1,0.1)$. It suggests that the iteration scheme breaks down from the first step and cannot compute $\tilde{D}_{j\geq 1}$ and the subsequent $\Delta G_{j\geq 1}$ as claimed in \cite{Ao2004}.
\begin{figure}[h]
	\centering
	\includegraphics[scale=0.5]{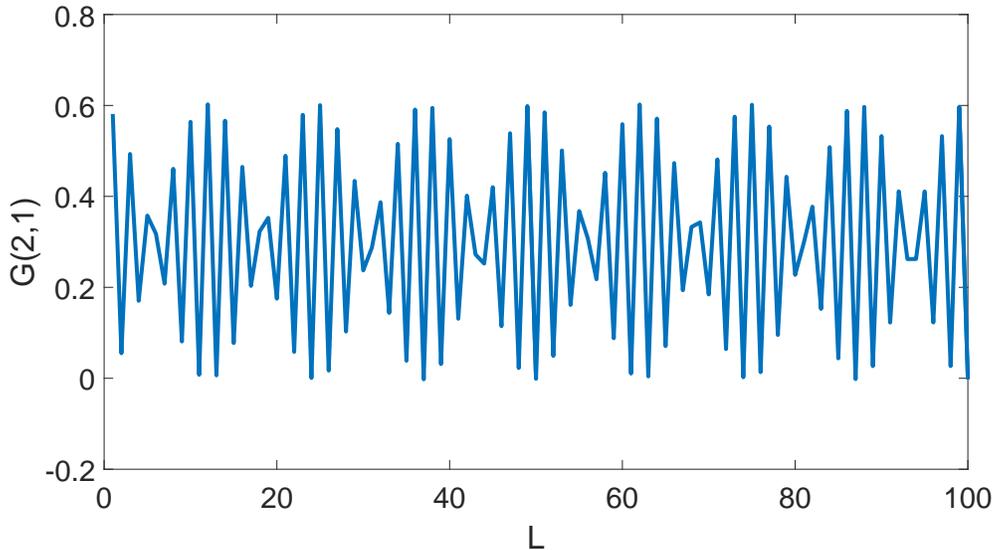}
	\caption{The divergence of gradient expansion scheme when computing $\Delta G_{0}$. $Y$-label is the value of partial sum to approximate $(2,1)$ element of $\Delta G_0$ at point $(0.1,0.1,0.1)^{\tau}$, and $X$-label is the upper bound of the summation. We find that the infinite series do not converge, suggesting the failure of gradient method to the simple problem  (\ref{example3d:sde}).}
	\label{fig:diverge}
\end{figure}
Hence, it is invalid to use gradient expansion as an argument towards uniqueness in Ao's SDE decomposition theory.

\end{itemize}

\section{Further Discussions}
We will discuss and clarify some other points raised in the YTA paper.
\begin{enumerate}
\item The speculation on the incorrectness of the proof in the ZL paper.
		
		In the YTA paper, the authors state that ``their non-uniqueness speculation is not only un-rigorous, it is also incorrect" in their final sentence. This is not serious. As we went over the YTA paper again and again, they do not show where the proof error is in ZL paper. It is not reasonable that the whole proof is correct in every place but it is incorrect finally! {\it Can the authors explicitly show which step  is incorrect in the proof of non-uniqueness in ZL paper?} We hope this is not about their ``proper'' boundary conditions.
		
		\item The SDE decomposition issue in the manifold case. 
		
It is true that $\phi^{AO}(x)=Constant$ on the manifold $S^1$, which can be derived from the result $\phi'(x)=0$ already shown in ZL paper. But the real issue lies in that the SDE decomposition is not well-defined in this case since one encounters $0=0$ situation.  The comment dodges this problem.  On the other hand, it is also stated that {\it  the potential $\phi(x)=-x$ in the winding number representation}.  We wonder how this solution satisfies the restrictions on the circle. The detailed setup of this statement is also not clearly mentioned.

				\item The uniqueness of Hamilton-Jacobi Equation issue.
		
		The YTA paper states ``the general reasoning for uniqueness was also used by ZL for the solution of the Hamilton-Jacobi equation for the potential function''. This speculation mis-understood the relevant part in ZL's paper.
		
		On the one hand, ZL paper does not make  general statement on the uniqueness of Hamilton-Jacobi Equation (HJE) and there is no need to do so since the FW quasi-potential is defined through the minimum of action functional instead of the solution of HJE. In general, the solution of the HJE for the quasi-potential is not unique. However, there has been serious mathematical study on the existence and uniqueness of the vanishing viscosity solution of the HJE and their numerical computations \cite{Lions, cameron}. To make the discussion simpler, ZL paper regards that the FW quasi-potential and potential function in Ao's SDE decomposition theory coincide with each other by default. This is based on the following two arguments. Firstly, the ZL paper discovers that \textit{any} solution of HJE (including the FW quasi-potential) can be  reinterpreted as Ao's potential function. Secondly, when there is only one stable point in the system, explicit conditions and rigorous proofs are available to guarantee that certain nontrivial solution of HJE must be the FW quasi-potential (for instance, Theorem 3.1 and related discussions in the book \cite{fw}). But in the more complex cases (e.g. multi-stable systems), the selection among multiple solutions of the HJE as the potential function is not straightforward, and the original SDE decomposition theory does not even recognize this issue. In fact, the SDE decomposition theory will suffer from more under-determinacy if the choice of potential function is not specified in such circumstance. The ZL paper relates Ao's proposal to the FW quasi-potential and large deviation theory, which provides more solid mathematical basis for the SDE decomposition. 
				
	   On the other hand, even if the above under-determinacy of the solution of the HJE is fixed,  there is urgent need to supply suitable conditions to fix the gauge in solving the PDE systems in SDE composition theory. In ZL paper it is found that if one chooses FW quasi-potential as the potential function, the reconstructed decomposition matrixes ($S,A,Q$) are generally not unique. While the SDE decomposition requires to compute the matrices first, and then obtain the potential function, this is far from straightforward intuitions. We need more explicit and more effective constructions beyond the so-called ``gradient expansion method'', at least excluding the degree of freedom of those reconstructed matrices.
		
		\item YTA also mentioned some other minor points as 6 items in their paper. As we see, the points $i), ii), iii)$ are not important. But the rest 3 points deserve comments.
		
		\begin{itemize}
			\item The terminology issue. It is not difficult to show that the condition $\hat{\sigma}(x)\hat{\sigma}(x)^{t}=2\varepsilon S(x)$  is equivalent to Eq.(1b) in the comment under the assumptions of SDE decomposition framework as reviewed in \cite{Ao2012}, which reduces to the Einstein relation in the detailed balance and constant $\gamma$ condition. The terminology ``generalization of Einstein relation" in ZL is natural for $\hat{\sigma}(x)\hat{\sigma}(x)^{t}=2\varepsilon S(x)$. The statement ``ZL erroneously referred $\cdots$" is not fair.
			
			\item The {\it stochastic integral  interpretation} of the zero mass limit issue. Indeed, similar limit has been studied in physics and applied mathematics \cite{Sancho, Majda}. Finally one gets {\it some} Fokker-Planck PDE describing the evolution of probability density. In one dimension, this Fokker-Planck PDE can be interpreted from the right-most endpoints integral of the over-damped SDE. But this interpretation is not valid in high dimensions. The extension of the over-damped system into a double dimensional system and then taking limit is not the usual definition of {\it stochastic integral interpretation  in mathematics}. The authors themselves may call it the microscopic interpretation.  But it is not fair to say ``It is an incorrect assertion $\cdots$" in ZL paper.
			
			\item The application to chemical jump process issue.
			
			Any reader can easily find that the successful application of large deviation and quasi-potential approach to the chemical jump processes in \cite{Li1,Li2, Li3} and other papers. It is also instructive to compare the methodology in \cite{AoJump} (stated by the YTA paper) with the approach in \cite{Li1,Li2, Li3} to see which is more fruitful.
			\end{itemize}
\end{enumerate}

Overall, the arguments in \cite{yta} about the ``incorrectness'' of ZL paper  does not hold. Before they claim the uniqueness result in the YTA paper, more considerations are required and more rigorous and detailed mathematical studies are demanded.

\begin{acknowledgements}
We thank Ping Ao, Hao Ge and Hong Qian for helpful discussions.
\end{acknowledgements}


\begin{thebibliography}{1}
		\bibliographystyle{unsrt}
		\bibitem{ZL}
		P. Zhou and T. Li, 
		\newblock{\em Construction of the landscape for multi-stable systems: Potential landscape, quasi-potential, A-type integral and beyond}, J. Chem. Phys, 2016, 144(9):553-558.
		
		\bibitem{yta}
		R. Yuan, Y. Tang and P. Ao,
		\newblock{\em On Uniqueness of" SDE Decomposition" in A-type Stochastic Integration}, arXiv:1603.07927v1, 2016.
		
		\bibitem{pnas}
		Kwon C, Ao P, Thouless D J. 
		\newblock{\em Structure of stochastic dynamics near fixed points}, Proc. Natl.  Acad. Sci. USA, 2005, 102(37):13029-33.
		
		\bibitem{Ao2004}
		P. Ao,
		\newblock{\em Potential in stochastic differential equations: novel construction}, J. Phys. A: Math. Gen.  37 (2004), L25.
		
		\bibitem{Ao2005}
		P. Ao,
		\newblock{\em Laws in Darwinian evolutionary theory}, Phys. Life Rev. 2 (2005), 117.
		
		\bibitem{Ao2008}
		P. Ao,
		\newblock{\em Emerging of stochastic dynamical equalities and steady state thermodynamics from Darwinian dynamics}, Commun. Theor. Phys. 49 (2008), 1073.
		
		\bibitem{Ao2009}
		P. Ao,
		\newblock{\em Global view of bionetwork dynamics: adaptive landscape}, J. Genet. Genomics 36 (2009), 63.
		
		\bibitem{Ao2012}
		R. Yuan and P. Ao,
		\newblock{\em Beyond Ito versus Stratonovich}, J. Stat. Mech. (2012), P07010.
		
		\bibitem{qian}
		Qian H, 
		\newblock{\em The Zeroth Law of Thermodynamics and Volume-Preserving Conservative Dynamics with Equilibrium Stochastic Damping}, Phys. Lett. A, 2012, 378(7-8):609-616.
		
		\bibitem{Sancho}
		J.M. Sancho, M. San Miguel and D. Durr,
		\newblock{\em Adiabatic elimination for systems of Brownian particles with
			non-constant damping coefficients},  J. Stat. Phys. 28 (1982), 291.
		
		\bibitem{Majda}
		P. Kramer and A. Majda,
		\newblock{\em Stochastic Mode Reduction for Particle-Based Simulation Methods for Complex Microfluid Systems}, SIAM J. Appl. Math. 64 (2003), 401.
		
		\bibitem{Li1}
		C. Lv, X. Li, F. Li and T. Li,
		\newblock{\em Constructing the energy landscape for genetic switching system driven by intrinsic noise}, PLoS One 9 (2014), e88167.
		
		\bibitem{Li2}
		C. Lv, X. Li, F. Li and T. Li,
		\newblock{\em Energy landscape reveals that the budding yeast cell cycle is a robust and adaptive multi-stage process}, PLoS Comput. Biol.  11 (2015), e1004156.
		
		\bibitem{Li3}
		T. Li and F. Lin,
		\newblock{\em Two-scale large deviations for chemical reaction kinetics through second quantization path integral}, J. Phys. A: Math. Theor. 49 (2016), 135204.
		
		\bibitem{AoJump}
		P. Ao, T.Q. Chen and J.H. Shi,
		\newblock{\em Dynamical decomposition of Markov processes without detailed balance}, Chin. Phys. Lett. 30 (2013), 070201.
		
		\bibitem{Lions}
		M.G. Crandall and P.L. Lions,
		\newblock{\em Viscosity solutions of Hamilton-Jacobi Equations}, Trans. Amer. Math. Soc. 277 (1983): 1-42.

		\bibitem{cameron}
		M.K. Cameron,
		\newblock{\em Finding the quasipotential for nongradient SDEs}, Phys. D, 2012, 241(18): 1532-1550.
		
		\bibitem{fw}
		M.I. Freidlin and A.D. Wentzell,
		\newblock{\em Random perturbations of dynamical systems}, Springer Science \& Business Media, 2012.
		
	
		

	\end{thebibliography}
\end{document}